\documentclass{PoS}
\usepackage[authoryear,square]{natbib}
\usepackage{epstopdf}

\bibpunct{(}{)}{;}{a}{}{,}

\title{Using SKA Rotation Measures to Reveal the Mysteries of the Magnetised Universe}

\ShortTitle{Using Rotation Measures to Reveal the Mysteries of the Magnetised Universe }

\author{
\speaker{Melanie Johnston-Hollitt}$^1$, 
Federica Govoni$^2$,
Rainer Beck$^3$,
Siamak Dehghan$^1$,
Luke Pratley$^1$,
Takuya Akahori$^{4,5}$,
George Heald$^{6}$,
Ivan Agudo$^7$,
Annalisa Bonafede$^8$,
Ettore Carretti$^{2,9}$,
Tracy Clarke$^{10}$,
Sergio Colafrancesco$^{11}$,
Torsten En{\ss}lin$^{12}$,
Luigina Feretti$^{13}$,
Bryan Gaensler$^{14}$,
Marijke Haverkorn$^{15,16}$,
Sui Ann Mao$^3$,
Niels Oppermann$^{17}$,
Lawrence Rudnick$^{18}$,
Anna Scaife$^{19}$,
Dominic Schnitzeler$^3$,
Jeroen Stil$^{20}$,
A. Russ Taylor$^{21,22}$,
and Valentina Vacca$^{12}$
\\ 
$^1$School of Chemical and Physical Sciences, Victoria University of Wellington, P.O. Box 600, Wellington 6140;
$^2$INAF - Osservatorio Astronomico di Cagliari,
Italy;
$^3$Max-Planck-Institut für Radioastronomie, 
Bonn, Germany;
$^4$Graduate School of Science and Engineering, Kagoshima University, 
Japan;
$^5$SKA Organization, Cheshire, UK;
$^6$ASTRON, Dwingeloo, The Netherlands;
$^7$Instituto de Astrofisica de Andalucia-CSIC, Granada, Spain
$^8$Hamburger Sternwarte, Universität Hamburg, 
Germany;
$^9$CSIRO Astronomy and Space Science, 
Australia;
$^{10}$Naval Research Laboratory, 
USA;
$^{11}$School of Physics, University of the Witwatersrand, South Africa;
$^{12}$Max Planck Institute for Astrophysics, 
Garching, Germany;
$^{13}$INAF Istituto di Radioastronomia, 
Bologna, Italy;
$^{14}$Dunlap Institute, University of Toronto, Canada;
$^{15}$Radboud University, the Netherlands;
$^{16}$Leiden University, the Netherlands;
$^{17}$Canadian Institute for Theoretical Astrophysics, University of Toronto, Canada;
$^{18}$Institute for Astrophysics University of Minnesota, 
USA;
$^{19}$The University of Manchester, 
UK;
$^{20}$The University of Calgary, Calgary, 
Canada;
$^{21}$University of Cape Town, South Africa;
$^{22}$University of the Western Cape, South Africa

\\
E-mail: \email{Melanie.Johnston-Hollitt at vuw.ac.nz}
}

\abstract{
We know that magnetic fields are pervasive across all scales in the Universe and over all of cosmic time and yet our understanding of many of the properties of magnetic fields is still limited. We do not yet know when, where or how the first magnetic fields in the Universe were formed, nor do we fully understand their role in fundamental processes such as galaxy formation or cosmic ray acceleration or how they influence the evolution of astrophysical objects. The greatest challenge to addressing these issues has been a lack of deep, broad bandwidth polarimetric data over large areas of the sky. The Square Kilometre Array will radically improve this situation via an all-sky polarisation survey that delivers both high quality polarisation imaging in combination with observations of 7-14 million extragalactic rotation measures. Here we summarise how this survey will improve our understanding of a range of astrophysical phenomena on scales from individual Galactic objects to the cosmic web.

}

\FullConference{
Advancing Astrophysics with the Square Kilometre Array\\
June 8-13, 2014\\
Giardini Naxos, Sicily, Italy}

\newcommand{\skipthis}[1]{}

\begin{document}

\section{Magnetism Science}

Magnetism science can be considered on a variety of scales and the origin and evolution of magnetic fields in the Universe is one of the great outstanding mysteries of modern astrophysics. We do not yet know how magnetic fields first arose in the Universe, nor if these fields formed via a top down or bottom up process. Although we are aware that magnetic fields manifest in a range of astrophysical objects, over vastly differing physical scales from the Mpc-scale halos in galaxy clusters to sub-pc scale jet features in radio galaxies, we are still speculating as to how those fields are important in the ongoing life-cycles of these objects. Additionally, understanding magnetic fields and their associated signatures in radiation fields is of crucial importance for experiments to detect and characterise the Epoch of Reionisation, and for models of inflation theory using B-mode polarisation of the cosmic microwave background radiation. The detailed properties of cosmic magnetic fields are fundamental to resolving the enduring mystery of the origin of ultra-high energy cosmic rays, and the Square Kilometre Array (SKA) will provide this information. Finally, wide-area surveys have the potential to unveil the magnetic field of the cosmic web itself. 

Although there are several methods for observing magnetic fields in other parts of the electromagnetic spectrum (e.g. optical polarisation, IR polarisation, observations of synchrotron emission in the optical and X-ray) or indirectly inferring the presence of magnetic fields (e.g.\ observations of magnetic Kelvin-Helmholtz or Parker instabilities \citep{Vikhlinin01} or Zeeman splitting \citep{Robishaw15}), all of these methods are limited in both their accuracy and range. Fortunately, observations in the radio are able to detect and characterise magnetic fields to high precision across much of the history of the Universe. 

Historically, however, it has been difficult to undertake detailed studies of the magnetic fields of large numbers of objects due to limits in sensitivity, resolution and difficulties with polarisation calibration of the current generation of radio telescopes. Additionally, understanding the complex nature of polarised signals was difficult prior to the advent of techniques to allow us to probe Faraday spectra generated by rotation measure synthesis and to statistically disentangle effects of the line-of-sight foregrounds. With the current generation of new or upgraded instruments, along with better analysis techniques and improvements in computer modelling, we are starting to see an increase in the type and quantity of polarisation science that can be undertaken. Polarisation surveys on current instruments such as the Polarisation Sky Survey of the Universe's Magnetism (POSSUM) \citep{Gaensler10} on ASKAP \citep{Hotan14}, the polarisation analysis of the GaLactic and Extragalactic All-sky MWA (GLEAM) survey \citep{Wayth15} on the MWA \citep{Tingay13} and the Multifrequency Snapshot Sky Survey (MSSS) (Heald et al., in prep.) on LOFAR \citep{vanHaarlem13} as well as smaller experiments on such instruments \citep{Bernardi13,Jelic14} will lay the foundations for an unprecedented period of magnetism science. The first phase of the SKA will take this work one step further providing instruments with high sensitivity, large bandwidths and good sky coverage, moving us into the  {\it "Era of Precision Magnetism Science"}. 

In order to exploit the vast potential of the SKA for magnetism science, the greatest priority for the magnetism community is a polarisation survey of the entire observable sky that provides both the polarisation properties of objects and the associated rotation measures. In fact, the goal of conducting an all-sky polarimetric survey on the Square Kilometre Array has been a mainstay for cosmic magnetism science for over a decade \citep{Gaensler04,Feretti04}. The resultant rotation measure (RM) grid continues to provide a compelling observational target to understand both the polarised sources themselves, and as a means to statistically probe numerous important extended foreground sources including the Milky Way, Magellanic Clouds, clusters of galaxies, the lobes of giant radio galaxies, external galaxies and perhaps even the elusive warm-hot intergalactic medium (WHIM). A suitably dense RM grid will provide information along the entire line of sight which, combined with improved statistical techniques, will allow us to answer such outstanding science questions as:\\
\begin{itemize}
\item What is the mechanism to generate and sustain magnetic fields in the Milky Way, Magellanic Clouds and other nearby galaxies?

\item How do magnetic fields influence galaxy formation and evolution?

\item How do magnetic fields manifest in HII regions, supernova remnants, planetary nebulae and high velocity clouds in the Milky Way?

\item Over what scales and at what strengths are magnetic fields generated in galaxy clusters and how does this correlate with cluster dynamical state?

\item What is the large-scale structure of the magnetised Universe and can we statistically detect the magnetic fields in the cosmic web?

\item What is the evolution of magnetic fields in intervening galaxies and clusters over cosmic time?
\end{itemize}

Here we will explore the science to be derived from a polarisation survey and the associated RM grid on the Phase 1 SKA and discuss the survey parameters needed to achieve characterisation of the most scientifically interesting objects.

\section{Polarisation Science}
In this section we will summarise the polarisation science priorities that have been identified for the SKA. We have grouped the topics in terms of the physical scale:
\begin{itemize}
\item Large scales ($\sim$Mpc): galaxy clusters and the cosmic web, 
\item Intermediate scales ($\leq$kpc): the Milky Way and nearby galaxies, and 
\item Small scales (kpc to pc): pulsars, masers, supernova remnants, high velocity clouds, shells and bubbles in the Milky Way and other galaxies. 
\end{itemize}

\subsection{Galaxy Clusters and the Cosmic Web}

\begin{figure*}
\centering
{\includegraphics[width=6.0in, trim= 0 0 0 0, clip=true]{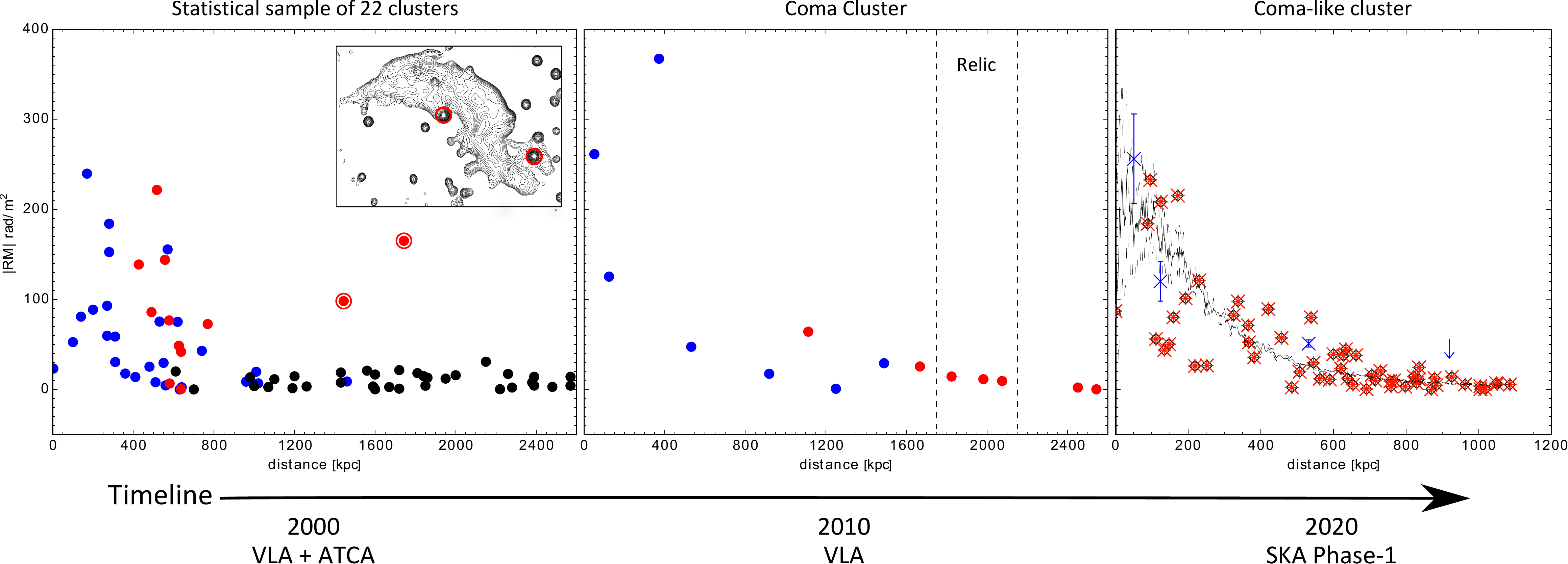}}
\vspace{-0.5cm}\caption{Statistical observations of RMs through galaxy clusters; left panel - observations of 22 clusters with 1-2 RMs per cluster, blue and black points are from \cite{Clarke01} and red points are from \cite{MJHetal04}. The plot includes the first observations through a radio relic showing an enhanced RM denoted by the two circled red points on the plot and in the inset showing their location in the relic \citep{MJH04a}; middle panel - observations of RMs through a single cluster, Coma \citep{Bonafede10,Bonafede13} the region of the Coma relic is shown by dashed lines; right panel - predicted number of RMs through a Coma-like cluster as seen with the SKA in Phase 1 \citep{Bonafede15}. The arrow at the bottom indicates the improvements to this type of work over time going from having only a statistical sample over many clusters 15 years ago to over 10 RMs per cluster in 2010 to finally several tens of RMs per cluster in the SKA era. Note the Galactic RM contribution has been estimated and removed in data presented in the left and middle panel using the latest results from \cite{Oppermann12}. }
\label{fig:limit}
\end{figure*}

We know there are magnetic fields permeating galaxy clusters on Mpc-scales via the presence of diffuse synchrotron emission seen as either central radio halos or peripheral radio relics, and via the statistical increase in rotation measures seen through cluster lines of sight \citep{Hennessy89,Kim91,Clarke01,MJH03,MJHetal04,Bonafede13}. However, to date, we lack detailed information on the extent or filling factors of the magnetic fields and their relationship to or influence on the dynamical state of the cluster. The SKA has long been thought to be an excellent tool for improving our understanding of clusters using statistical RMs \citep{Feretti04}, and more recently it has been shown to allow the observation of polarised filaments in cluster halos across a range of redshifts \citep{Govoni13,Govoni15}. 

\begin{figure*}
\centering
{\includegraphics[width=4.0in, trim= 0 0 0 0, clip=true]{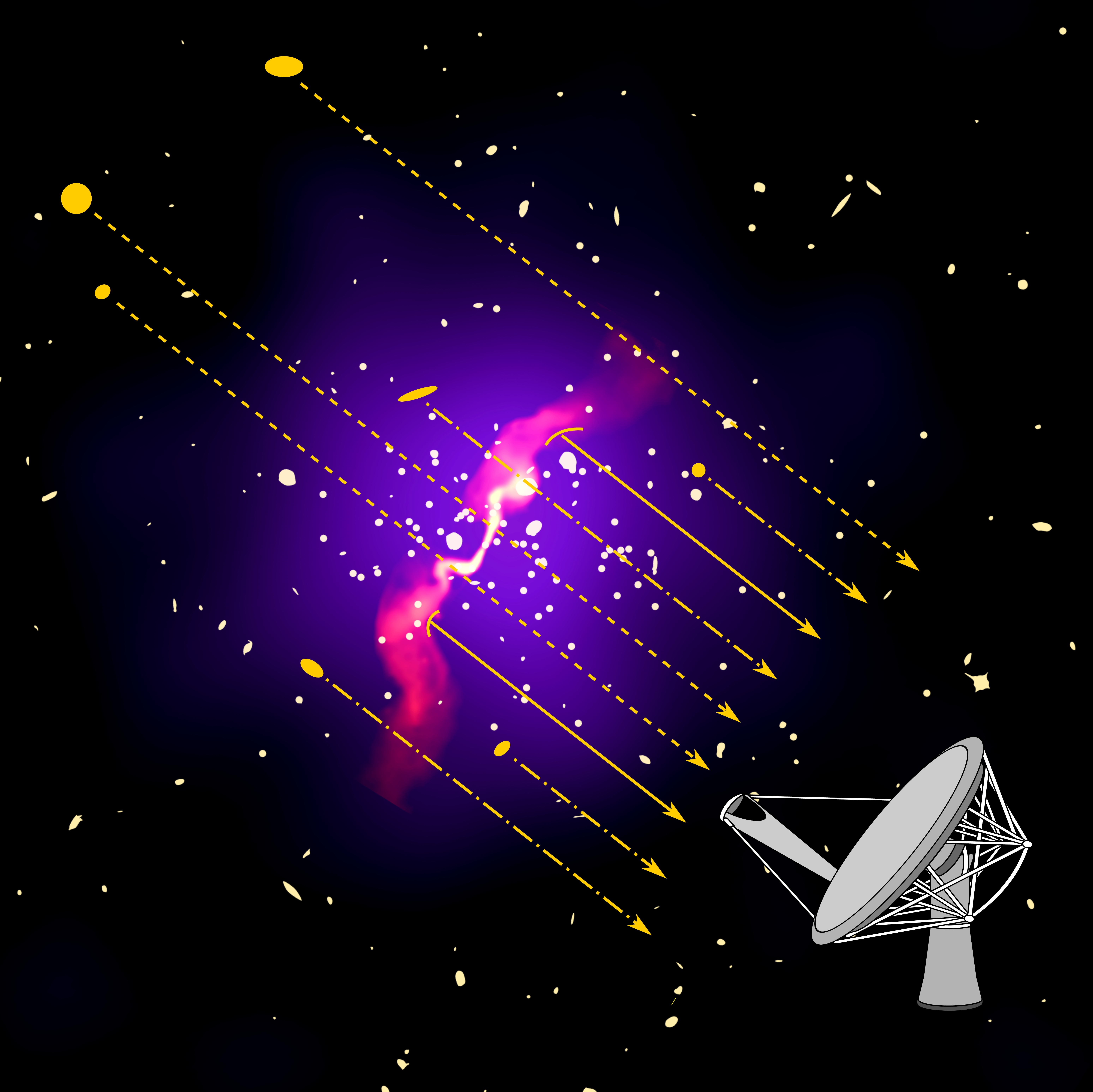}}
\vspace{-0.3cm}\caption{Schematic of a nearby galaxy cluster showing X-ray emission in purple, an extended radio source in pink, and unresolved radio sources in white if unpolarised and gold if polarised. Different path lengths to polarised sources are marked including unresolved background radio galaxies (dashed lines), unresolved embedded sources (dot-dashed lines) and extended embedded sources such as large tailed radio galaxies, the lobes of which are polarised and can be used as screen to examine the cluster magnetic field (solid lines). The wealth of sources located at different locations within the ICM will allow the first Faraday tomography of the magnetic field in galaxy clusters.}
\label{fig:limit}
\end{figure*}

Progress in our understanding of cluster magnetic fields has been hampered by lack of detail regarding the contributions from foregrounds (both intrinsic and extrinsic), poor sampling of background sources for RM measurements (a result of lack of sensitivity) and poorly sampled polarisation data leading to incorrect RM fits (the result of lack of bandwidth). Progress to statistically sample cluster magnetic fields via the use of RMs of background sources has been slow but steady. We have moved from having to amass statistical values over a number of clusters in which there are only 1-2 measurements per cluster \citep{Clarke01,MJHetal04} to being able to easily detect several sources in an individual cluster \citep{Bonafede13}. Predictions for SKA1 show we will have several tens to a few hundreds of background sources to probe clusters up to a redshift of 0.1\footnote{Assuming a 1 Mpc radius and extrapolating the current polarised source counts to 4 $\mu$Jy \citep{Rudnick14,Hales14,Stil14} gives 60-115 RMs per cluster at z=0.05 and 15-30 at z=0.1, very large, nearby clusters like A3266 would have 180-360 background RMs.}. Additionally, we will be able to conduct statistical studies with 1-2 RMs per cluster out to a redshift of 0.5, this is important to understand the way magnetic fields grow in the intracluster medium (ICM). The vast increase in the number of RMs available in the outskirts of clusters will also allow measurements through magnetic fields in the cluster relics that are thought to be generated by shocks. Observations have shown relics to have highly aligned magnetic fields running perpendicular to the direction of shock propagation. To date only observations through the NW relic in A3667 have shown an increase in RM \citep{MJH04a}, but simulations predict such enhancements will be readily detectable, particularly with SKA2 \citep{Bonafede15}. Observations of relic RMs combined with RMs of sources seen in projection through relics will greatly assist in disentangling the shock geometry and allow the first statistical samples of the magnetic fields in shocks to be constructed. Figure 1 provides a schematic to illustrate the improvement in background RMs seen through cluster magnetic fields commencing in 2000 and going through to the expected SKA levels in 2020. By allowing detailed measurements of the magnetic field strength and distribution of individual galaxy clusters, the door is opened to examine changes in the magnetic field as a function of other cluster properties such as X-ray luminosity/mass, dynamical state and importantly will allow us to determine the radial profile of the field and its connection to the gas density. 

For nearby galaxy clusters (z $\leq$ 0.1) we will have sufficient sources in both the background and embedded within the ICM to perform Faraday tomography of the magnetic field of the cluster. In particular, we will have a host of background sources, embedded point sources and embedded extended sources such as tailed radio galaxies with which to probe different path lengths through the ICM (see Figure 2). The use of embedded cluster sources to probe the magnetic field has already occurred in a few isolated examples \citep{Guidetti08,Pizzo11,Pratley13}, but to date we have been unable to undertake the detailed tomographic studies the SKA will allow. In fact, not only will the SKA allow Faraday tomography of the field via the use of a large number of sources as probes to the cluster RM, the resolved structures in some tailed radio galaxies such as the so-called `corkscrew' radio galaxies could potentially provide information on the magnetic field in clusters over very small scales \citep{MJH15a,MJH15b}. 

Moving away from statistical studies, the SKA will also allow detailed mapping of the magnetic fields associated with cluster relics and potentially even halos which are expected to be detected in large numbers \citep{Cassano15}. Whilst we have seen polarised filamentary structure in only two cluster halos \citep{Govoni05,Bonafede09}, we expect the SKA, and in particular SKA2, to have sufficient sensitivity to image such structures \citep{Govoni13,Govoni15}, potentially at a range of redshifts\footnote{The ability to detect polarised emission in filaments is a function of the resolution and sensitivity, for SKA1 we expect to be able to see polarisation from only the most luminous radio halos (L$_{1.4} \geq$ 3 $\times 10^{25}$ W/Hz), while SKA2 will be able to achieve similar results for intermediate brightness halos (L$_{1.4} \geq$ 2 $\times 10^{24}$ W/Hz) see Govoni et al.\ (2015) for further details.}. This provides the exciting possibility to measure the evolution of central cluster fields as a function of time, examining variables such as changing strength, degree of order, turbulence scales and power spectra. Magnetic fields in galaxy clusters are thought to be turbulent and thus will produce characteristic Faraday depth spectra observable with sufficiently high resolution in Faraday depth space, using the SKA an examination of clusters will thus be able to disentangle different spectral templates attacking questions about field geometries, in particular questions surrounding the existence of asymmetry and vertical components to the field. 

Looking on the largest scales, one of the most exciting challenges for the SKA is to discover and characterize the WHIM in the cosmic web, which is the last, major unproven piece predicted by standard inflationary cosmology. Since it is likely that the intergalactic magnetic field (IGMF) permeates the WHIM in the cosmic web \citep{Ryu12}, in the coming decades the SKA will be the most promising facility to discover and characterize such fields. 
Direct imaging of the cosmic web has been shown to finally be a possibility with SKA1 \citep{Vazza15,Giovannini15} and it will also be possible to measure the dispersion measure (DM) of the WHIM via fast radio bursts \citep{Macquart15}, however, we also expect to detect the statistical RM signal \citep{Taylor15}. Previous attempts to detect the statistical RM have been difficult due to foreground contamination \citep{Schnitzeler10,Xu14}, but the dense RM grid obtained with the SKA dramatically improves the ability to account for intervening foregrounds and the broad bandwidths of the telescope provide an opportunity to select sources passing through the WHIM \citep{Akahori14b} based on Faraday depolarisation and Faraday tomography \citep{Gaensler15}. The detection of the IGMF and the resultant comparison with theoretical predictions \citep{Akahori14a} will be a breakthrough in our understanding of the way in which magnetic fields assemble on large-scales in the Universe.

{\it For further information see the following chapters in the 2015 SKA Science Case: \cite{Bonafede15,Cassano15,Gaensler15,Giovannini15,Govoni15,MJH15a,Macquart15,Taylor15,Vacca15,Vazza15}.}

\subsection{Magnetic Field of the Milky Way}
\begin{figure*}
\centering
{\includegraphics[width=5.7in, trim= 0 0 0 0, clip=true]{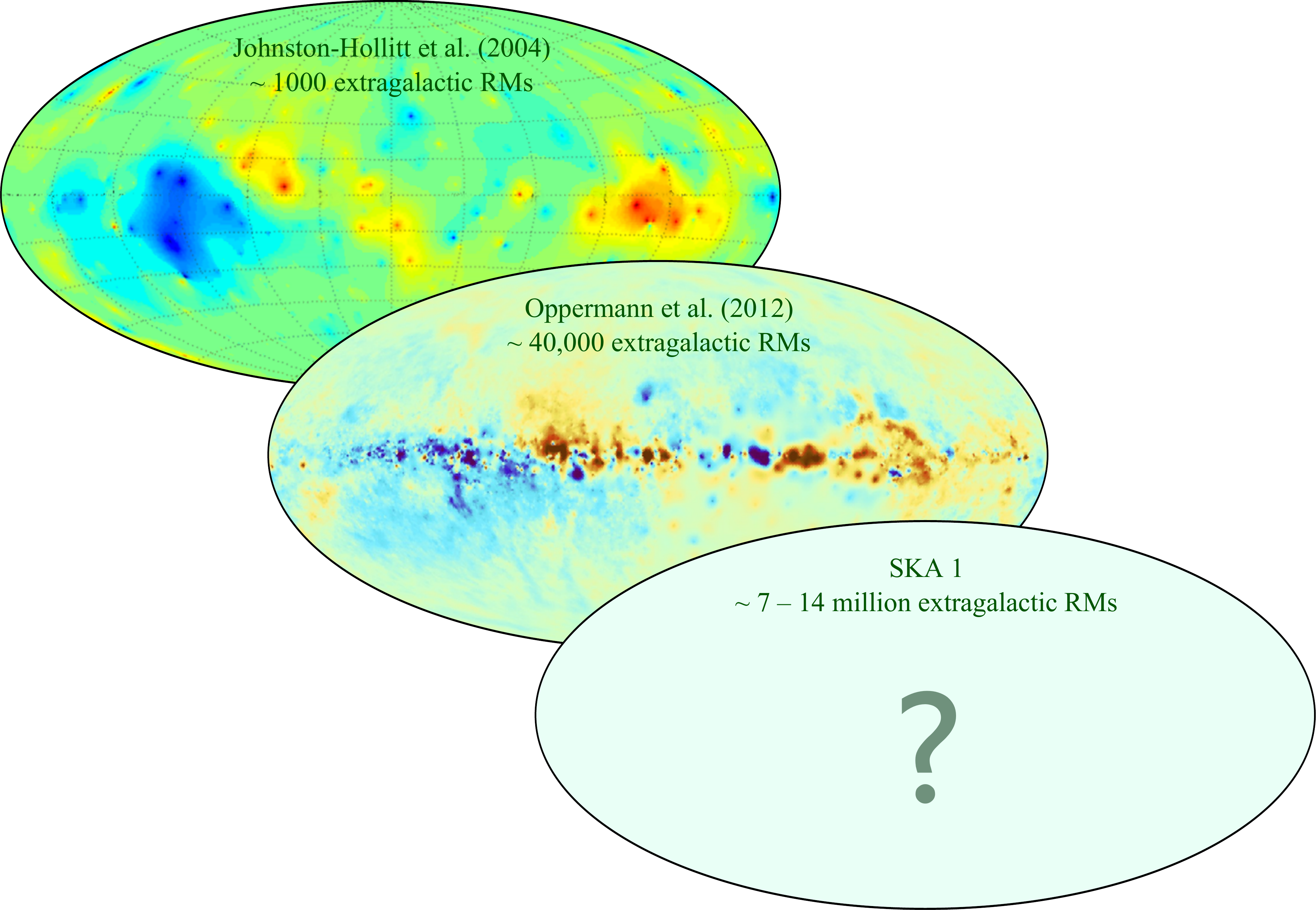}}
\vspace{-0cm}\caption{Top projection: The RM sky in Galactic coordinates as interpolated from $\sim$1000 extragalactic RMs over a decade ago \citep{MJH03,MJH04b}. Middle projection: The RM Sky as determined from more sophisticated signal processing methods for $\sim$40,000 extragalactic RMs \citep{Oppermann12,Oppermann15}. Note that the large-scale features of the field are largely unchanged between the top and middle panel, but the small scale information regarding the magnetic field of the Milky Way is greatly improved with a higher density of RMs. The bottom panel denotes that an all sky RM survey on SKA phase 1 with a sensitivity of 4 $\mu$Jy/beam at 2" resolution should provide 7-14 million extragalactic RMs with which to probe the RM sky. Red colour scales denote positive RMs and magnetic fields coming out of the plane of the sky, whilst blue colours denote negative RMs and fields going into the plane of the sky.}
\label{fig:limit}
\end{figure*}

Mapping the magnetic field of the Milky Way has been steadily improving over the last decade. The use of extragalactic background sources, embedded pulsars and observations of the diffuse synchrotron emission in polarisation surveys \citep{Reich04,Haverkorn06,Stutz14} have all played important roles in examining the large-scale magnetic field of our Galaxy \citep{Stil11,Oppermann12}. Such work continues to reveal surprising and previously unknown features such as giant magnetised outflows \citep{Carretti13}, and has permitted mapping of the magnetic field in a range of discrete Galactic objects \citep{McG10,LHS11,Purcell15} and given the first information on the Mach number of Galactic turbulence \citep{Hill08,Gaensler11}. Despite these advances, many questions as to the exact field configuration remain, including controversy surrounding the direction of the large-scale regular field in the spiral arms, details of the field in the Galactic centre and information on the power spectrum of the turbulent field. The SKA will resolve many of these questions and allow the first reconstruction of the 3D field via detailed Faraday tomography using external and embedded sources. 

The leap forward that SKA will provide in our attempts to reconstruct the magnetic field of the Milky Way cannot be overstated. In the last decade, we have moved from initial attempts to reconstruct the large-scale field using only $\sim$1000 RMs gathered from inhomogeneous observations combined with relatively simple interpolation \citep{MJH04b} to having access to $\sim$40,000 RMs \citep{Taylor09} and advanced statistical techniques \citep{Oppermann12,Oppermann15}. Improving the sampling rate by a factor of 40 to the level of 1 source per square degree has facilitated a tremendous leap forward for Galactic magnetism (see Figure 3), the phase 1 SKA will further improve the sampling rate by over two orders of magnitude, providing an astounding 7-14 million discrete RMs at a sampling rate of approximately 230 to 450 sources per square degree. Additionally, SKA1 is expected to increase the number of Galactic pulsars by a factor of 5 \citep{Keane15} taking the total to roughly 10,000. Parallax measurements will provide accurate distances to a large number of pulsars distributed through the Galaxy \citep{Smits11,Keane15,Janssen15}. As with observations of nearby galaxy clusters, combining the rotation measure information from extragalactic background sources and embedded pulsars at precisely known locations will allow an unprecedented 3D reconstruction of the magnetic field of the Milky Way and will resolve uncertainties around field reversals in the spiral arms and complexity in the field configuration in the Galactic Centre as well as provide detailed constraints on the power spectrum of the Galactic magnetic field \citep{Haverkorn15,Han15}.

{\it For further information see the following chapters in the 2015 SKA Science Case:  \cite{Haverkorn15,Han15}.}

\subsection{Magnetic Fields on kpc to pc star forming scales}

As noted above, the currently available extragalactic RMs have a density of roughly 1 source per square degree which has been sufficient to obtain information on the magnetic fields in a host of Galactic objects including supernova remnants, HII regions, high velocity clouds, star forming regions and shells and bubbles in the Milky Way. As we move to sampling at a rate of approximately 230 to 450 sources per square degree we have the opportunity to probe a huge number of discrete Galactic objects on kpc to pc scales in both the Milky Way and the Magellanic Clouds (see \cite{Haverkorn15} for further details).  Theoretical models suggest that magnetic fields play a vital role in the evolution of dense molecular clouds and the associated star formation within them, this is supported by the extraordinarily close radio--IR correlation observed locally within galaxies \citep{Tabatabaei13}. The generation of small-scale magnetic fields by supernova-induced turbulence is one possible scenario to explain this correlation, or it may be that a more fundamental relation exists between magnetic fields and molecular gas. Investigating these relations both within the Milky Way and nearby galaxies is crucial to understanding the role of magnetic fields in the structure formation of the molecular interstellar medium (ISM), and in particular, the formation
of giant molecular clouds that are the cradles of massive star formation. Observations of the {\it in-situ} synchrotron radiation emitted by dense molecular clouds is also a possibility with SKA \citep{Dickinson15} and imaging the polarised emission from such clouds will provide constraints on the internal magnetic fields and their distribution in these clouds, which will be crucial to addressing questions regarding the effect of magnetic fields in star formation. 

{\it For further information see the following chapters in the 2015 SKA Science Case: \cite{Dickinson15,Haverkorn15}.}

\subsection{Magnetic Fields in Nearby Galaxies}

\begin{figure*}
\centering
{\includegraphics[width=5.8in, trim= 0 0 0 0, clip=true]{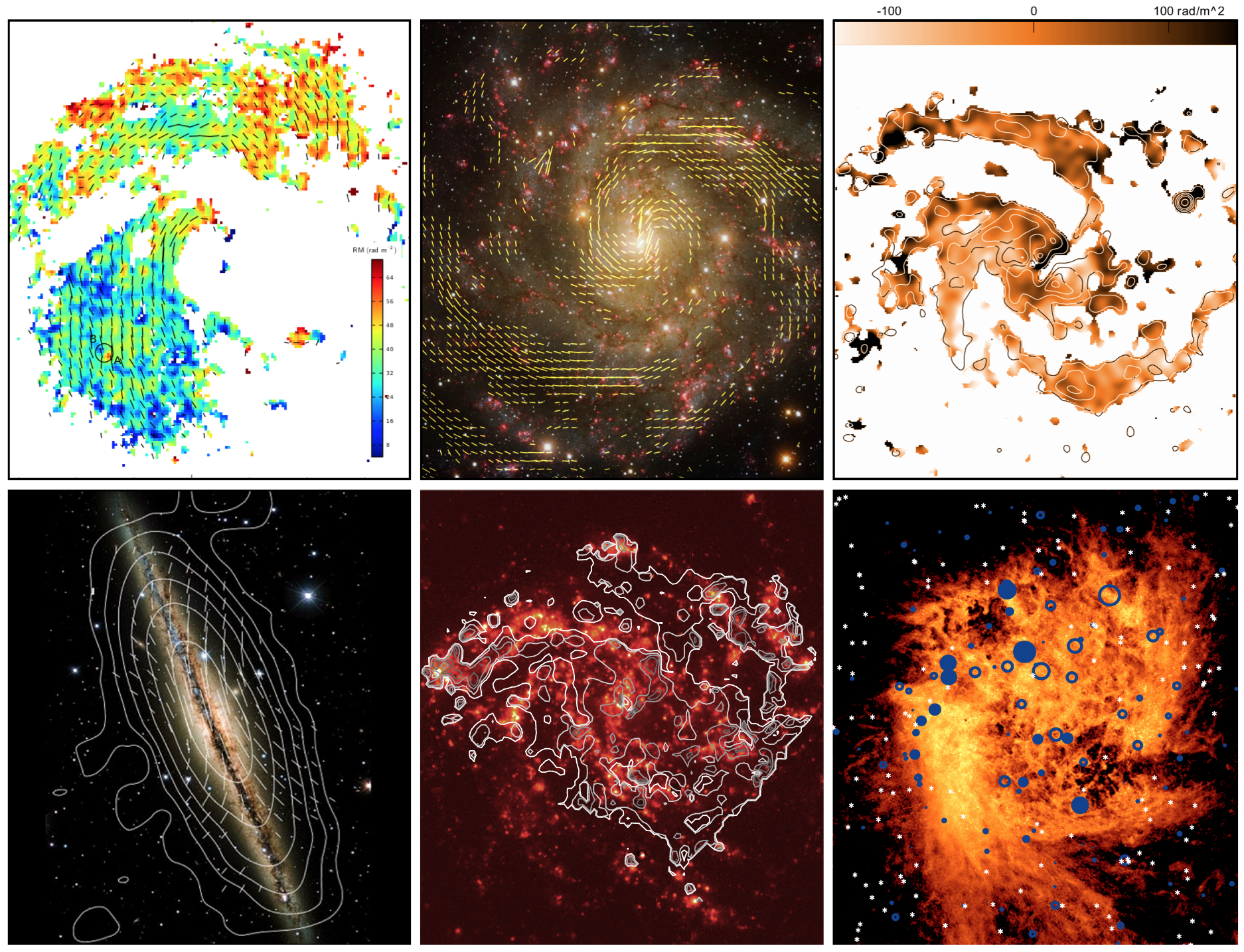}}
\vspace{-0cm}\caption{Examples of information that the proposed polarisation survey will provide on the magnetic field of nearby galaxies using different techniques. \emph{Top left:} Rotation measure values overlaid with polarisation vectors for  NGC 6946 \citep{Heald12}. \emph{Top middle:} Magnetic field lines from synchrotron polarisation observations overlaid on an optical image of IC342 \citep{Beck15b}. \emph{Top right:} rotation measures associated with NGC 6946 \citep{Beck07}. \emph{Bottom left:} Total radio emission and magnetic field vectors of the edge-on galaxy NGC 891 overlaid on the optical image from CFHT \citep{Krause09}.  \emph{Bottom middle:} Contours of the strength of Faraday dispersion for NGC 6946 overlaid on H$\alpha$ emission (Williams et al. in prep). \emph{Bottom right:} Extragalactic rotation measures seen in projection through the Large Magellanic Cloud \citep{Gaensler05}. Note the colour scales of some images have been altered from the original to better highlight features.}
\label{fig:limit}
\end{figure*}

Ordered magnetic fields of spiral shape are observed in all spiral galaxies and even
in flocculent galaxies without a spiral pattern in the gas \citep{Beck2015c}. Ordered magnetic fields are generally strongest
in the regions between the material spiral arms, in some cases forming ``magnetic arms'' with
exceptionally high degrees of polarisation \citep{Beck96}. Although models of large-scale (mean-field) dynamos can basically explain
these phenomena \citep{Chamandy15,Moss15}, the physics of galactic dynamos is still far from being understood.
Present-day MHD models (including dynamo action) can achieve resolutions of the physically relevant
scale of turbulence of a few tens of pc within kpc-size boxes of the ISM \citep{Gent13}. Global MHD models including
gravitational instabilities like bars and spiral arms, and MHD models of galaxy evolution are still to come.

In contrast with simplistic models, real galaxies have complex field patterns that can be described by a power spectrum of Fourier modes
imprinted in Faraday rotation measures. Field reversals, density waves, bars and mergers also
leave specific signatures in the power spectrum which SKA will be sensitive to. Field reversals between spiral arms on scales of several kpc,
like that observed in the Milky Way, have not been detected in external galaxies so far. Such reversals
could be present as relics from the early epoch of galaxy evolution, or triggered by a major merger. The number and
extent of field reversals may provide new insight into the history of galaxies. RM data
with high spatial resolution (less than about 100\,pc) are required for a systematic investigation of these phenomena.

The formation of spiral arms is another area where the role of magnetic fields is an open question. Density waves
including magnetic fields have two modes of propagation \citep{Lou98}. In the fast MHD mode, magnetic field waves that are
in-phase with the gas waves may stabilize spiral arms. The out-of-phase magnetic fields of slow MHD
waves may become unstable and develop into Parker loops which can be observed via azimuthally periodic variations
in Faraday rotation measures \citep{Beck15b}.

High-resolution polarisation images of nearby galaxies are needed to study and understand the physics of
large- and intermediate-scale interstellar magnetic fields, supplementing observations of local
small-scale magnetic fields in the Milky Way. There are three basic techniques that we can employ to explore the magnetic field in nearby galaxies.

The first and most straightforward approach is direct imaging of the polarised synchrotron emission emitted
by a galaxy to reveal the features in the large-scale ordered fields in the plane of the sky, preferably
at high frequencies (above 3\,GHz) where the angular resolution is highest and 
the effects of Faraday rotation and depolarisation are sufficiently small to observe the field's intrinsic orientation and
degree of order. 

A complementary approach at somewhat lower frequencies is to investigate the polarised emission and its
variation across a wide frequency band. The resulting Faraday rotation measure and frequency-dependent
depolarisation can be used to examine the detailed internal structure of the magnetoionic medium, both
across the galaxy and along the line-of-sight via Faraday tomography \citep{Heald15}. Details that can be studied include
the coherence length of the line-of sight magnetic field, the large-scale three-dimensional magnetic field
structure including the vertical structure of magnetic fields and the properties and dynamical impact of
magnetic fields in the disk-halo interface of star forming galaxies. To achieve
excellent resolution in Faraday depth, we require a wide coverage in $\lambda^2$, i.e.\ wide frequency
coverage in a lower frequency band (such as proposed for the polarimetric sky survey).

Finally, as with the Milky Way, we can use the RMs of polarised background galaxies to obtain information about
the magnetic field in the foreground galaxy, averaged along the line-of-sight (the ``RM grid'' approach).
Because this method does not depend on the occurrence of synchrotron-emitting cosmic-ray electrons (CREs) 
in the foreground galaxy, it can also be exploited to detect and study magnetic fields in the outer regions of
galaxies where CREs cannot reach, providing comprehensive information about the extent to which magnetic fields are expelled by outflows at these large radii. 

Figure 4 provides examples of each technique. These methods are highly complementary and provide us with
the unique opportunity to better understand the magnetic field in a range of galaxies. SKA will permit observations of 
nearby galaxies using these techniques at kpc resolution out to a redshift of at least 0.025, providing a comprehensive sample of all galaxy field geometries in the local volume.

{\it For further information see the following chapters in the 2015 SKA Science Case: \cite{Beck15,Heald15}.}

\subsection{Magnetic Fields in Active Galactic Nuclei and the Faint Polarised Sky}
\label{faint}
The lobes and jets of Active Galactic Nuclei (AGN) are rich laboratories to investigate a range of phenomena associated with magnetised plasmas. The polarisation survey on SKA1 will image and provide RMs for a vast number of AGN across cosmic time allowing a diverse set of studies including direct imaging of the magnetised lobes and jets of nearby galaxies, examination of the rotation measure structure \citep{Guidetti08,Guidetti12,O'Sullivan13} and depolarisation characteristics \citep{Fomalont89} across such lobes, use of the lobes themselves as Faraday screens to investigate their local environment, and use of background galaxies to probe structure within the lobes \citep{Feain09}. Maps such as the spectacular image of Fornax A shown in Figure 5, which shows the rich depolarisation structure evident across the radio lobes, will be possible for all nearby radio galaxies. Such data allow exploration of structure within the lobes themselves and detection of local depolarising screens (e.g.\ the `ant' in the western lobe of Fornax). A census of such emission as a function of cosmic time will provide constrains on the environment surrounding these sources \citep{Gaensler15,Taylor15}. 

\begin{figure*}
\centering
{\includegraphics[width=4.0in, trim= 0 0 0 0, clip=true]{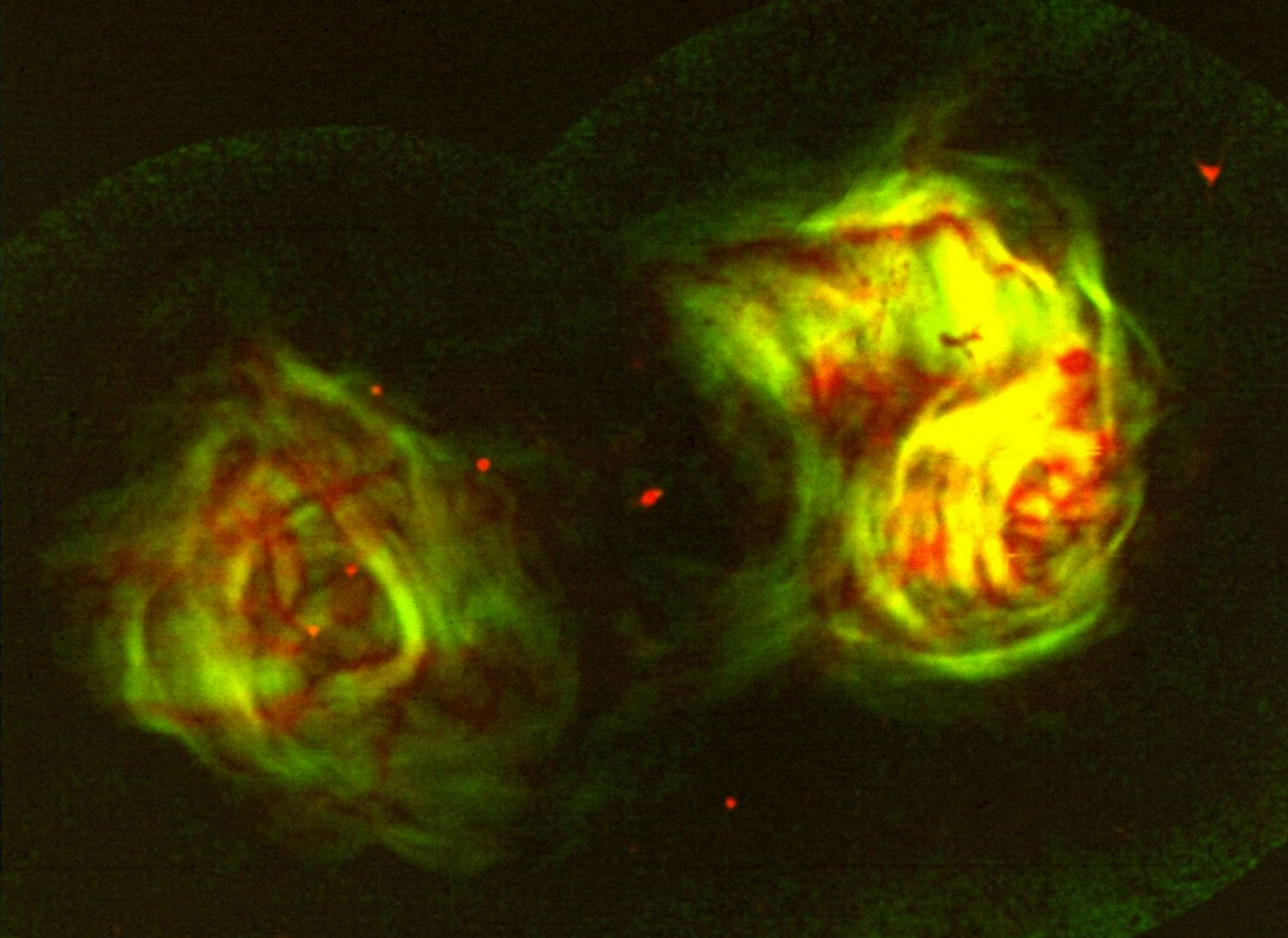}}
\vspace{-0cm}\caption{Fornax A shown here in a hue-intensity plot in which flux is represented by intensity and fractional polarisation is represented as hue with regions of high depolarisation appearing red. The depolarisation structures presented here result from a range of phenomena including internal depolarisation in the lobes, depolarisation due to foreground gas (e.g.\ the `ant' in the western lobe) or intervening galaxies and instrumental beam depolarisation \citep{Fomalont89}. SKA1 will routinely produce information across the resolved lobes of radio galaxies. }
\label{fig:limit}
\end{figure*}

The SKA will enable observations an order of magnitude deeper than current instruments and the observations will be carried out over a large enough area of the sky to provide a statistical overview of the magnetic properties of the faint radio source population. Deep imaging with SKA1 will probe the polarised properties of normal galaxies to high redshifts allowing us to monitor their evolution of these properties over cosmic time and to investigate when magnetic fields in galaxies first emerged and how quickly they were amplified. A deep survey capable of probing star forming galaxies over a wide range of redshifts will capture the population in different stages of evolution. In particular, we will observe galaxies as they convert gas to stars, a process which is believed to be strongly tied to their magnetic properties. 

Additionally, examining the polarisation properties of active galaxies over cosmic time provides a powerful tool to constrain jet magneto hydrodynamic and emission models. Deep SKA surveys will thus present the first opportunity to address the longstanding problem of the composition of AGN jets, and their plasma acceleration using large, well designed samples. With such a survey we will be in the ideal position to finally understand the effect of increasing fractional polarisation with decreasing flux density/observational frequency, an effect that suggests there must be either changes in the internal source field structure or their Faraday screens, likely related to the immediate environment of the galaxy. Furthermore, deep SKA surveys will probe the AGN population to redshifts of up to 10, capturing these sources at a time when the interstellar and intergalactic media were considerably denser than the present day. Observing these frustrated sources as they attempt to break out of their host galaxies provides an exciting and potentially unique opportunity to probe not only the magnetic fields, but also the surrounding intergalactic medium \citep{Gaensler15,Taylor15}. 

Finally, although we will get the first hints of the magnetic field in the cosmic web through the wide-field polarisation survey, the challenging nature of detecting such faint RMs will be most appropriately ameliorated via a deep survey at high Galactic latitudes, away from galaxy clusters \citep{Taylor15}. 

Even with the sensitivity and depth of the SKA there will be a significant population of objects for which no polarisation signal is detected. Recent work on the faint polarised source population has shown that information can still be obtained on the polarisation properties of these faint sources via stacking analysis \citep{Stil14}. In the context of the SKA stacking provides an extension to the results of deep surveys, particularly regarding understanding the fractional polarisation as a function of redshift and as a means to investigate the polarisation luminosity function \citep{Stil15}.

{\it For further information see the following chapters in the 2015 SKA Science Case: \cite{Agudo15,Gaensler15,Laing15,Peng15,Stil15,Taylor15}.}

\subsection{Other Magnetism Science}
Although the wide-area polarisation survey and the associated RM grid experiment cover a great deal of magnetism science, there are other important magnetism topics which are worth elaborating on here. Studies of the Zeeman effect in both emission and absorption towards atomic and molecular clouds in the Milky Way with SKA1 will provide information on the magnetic field in the warm and cold, neutral components of the Galaxy while Zeeman observations of OH masers provide information on large-scale galactic fields \citep{Robishaw15}. Thus all Zeeman observations will be highly complementary to the extensive Faraday rotation measure observations. Looking forward to SKA2 such work could potentially be extended to nearby galaxies and in particular observations of megamasers are expected to extend out to $z=1$.

Finally, one intriguing possibility is to use the SKA to probe the nature of dark matter in cosmic structures, from dwarf galaxies to galaxy clusters. Under the assumption that dark matter annihilation produces an observable component to the synchrotron emission detected in cluster and galaxy halos on top of that generated from cosmic rays and star formation, we should detect low-level polarised emission associated with these structures. This is especially the case for (sub-)structures with low star formation activity or cosmic ray density. Studying the magnetic fields in clusters and galaxies is therefore crucial to disentangling the dark matter particle density from the magnetic energy density that is contributing to the predicted annihilation-induced synchrotron emission. The RM grid experiment will be unique in these respects being able to provide simultaneous measurements of the magnetic field while putting constraints on the putative annihilation-generated radio halo emission \citep{Colafrancesco15}.

{\it For further information see the following chapters in the 2015 SKA Science Case: \cite{Colafrancesco15,Robishaw15}.}

\section{Polarisation Surveys}
In order to capitalise on the immense opportunity the SKA provides for magnetism science it is vital that a polarisation survey over the entire accessible sky at 950-1760 MHz (the so-called SKA1-MID band 2) is undertaken down to a sensitivity of 4 $\mu$Jy/beam. This survey must provide not only full polarisation imaging at 2" resolution, but also deliver the RM catalogue of the expected 7-14 million discrete extragalactic sources, or provide sufficiently calibrated data for these RMs to be determined in post-processing. Such a survey will be a powerful legacy dataset with which to undertake an extensive range of high impact science including the RM grid experiment, imaging of the magnetic field of all nearby galaxies within 100 Mpc radius, detailed mapping of the magnetic field structure in cluster relics and possibly halos, and finally the survey will probe the magnetic field in a range of astrophysical objects from the jets and lobes of radio galaxies to supernova remnants. This survey will thus advance a great deal of progress in a broad range of science cases and is the highest science priority of the SKA magnetism community. We estimate that the survey will require 2.5 years of observing time (inclusive of calibration overheads) on the SKA1-MID instrument in South Africa. This will produce a legacy survey of vast scientific potential, not only for magnetism, but for a host of high-priority continuum science (see \cite{Prandoni15} for further details). The polarisation survey is well suited to be undertaken in a commensal mode with the survey proposed by the SKA Continuum working group and with the Galactic HI absorption grid experiment. Additionally, there is strong synergy between the observations of nearby galaxies that will be undertaken as a matter of course in the wide-area polarisation and continuum survey, and the spectral line observations of nearby galaxies that will also be performed by the HI group. Comparisons of the polarisation, continuum and HI data for nearby galaxies is pertinent to understanding the physical connections between gas, star formation, and magnetic fields. 

Additionally, we propose a series of smaller, deeper surveys to examine the polarisation properties in the very faint galaxy population and to allow an expansion of the accessible range over which we can probe the evolution of magnetism in different classes of objects. The first of these surveys would be a high Galactic latitude survey of order 10 square degrees from 950-1760 MHz. Ideally such a survey would reach a depth of 75 nJy/beam at 2" resolution and would allow the first statistically significant investigation of the faint polarised sky (see Section \ref{faint} for further details). The definition of other deep survey areas will be the subject of future discussions.  Finally, we note that all continuum observations on the SKA will provide useful polarisation information and we would encourage the SKA to provide full polarisation products, including RMs for all observations.

Looking forward to SKA2 we will have even greater sensitivity, resolution and potentially larger bandwidths, all of which will allow another leap forward for magnetism science. We will continue to enhance the density of sources on the RM grid, sampling at least 40 million radio sources allowing an even finer level of detail with which to resolve magnetic fields in a diverse range of objects. The higher spatial resolution will take us into a resolution regime which is currently only accessible with VLBI, but unlike present day VLBI will not be restricted to bright, compact objects. This will be an almost unimaginable change in radio astronomy allowing for unprecedented imaging of jets of AGN, magnetic fields in nearby galaxies, and detection and imaging of the polarised component of radio halos out to redshifts close to 2 when the first clusters were formed. Additionally, we will delve deeper into the history of the Universe in our quest to resolve the overarching questions of where, when and how magnetic fields arose in the Universe.

\section{Summary}

Future wide-area radio surveys via instruments such as SKA1 will provide very sensitive and yet observationally affordable ways to explore a wide range of science goals, including those associated with the detection and characterisation of magnetic fields over a range of scales. The planned polarisation survey on SKA1 and its associated RM grid will give transformational science for almost every class of astronomical object. With this survey we will undertake a complete census of magnetic properties in normal galaxies and AGN as a function of cosmic time, probe the detailed physics of magnetised plasmas in both the jets of AGN and the cores of galaxy clusters, and measure RM enhancements in shock-driven relics out to a redshift of 0.5. We will conduct Faraday tomography on the magnetic fields in clusters, nearby galaxies and the Milky Way in astounding detail placing limits on the extent, turbulence and radial profile of these fields. Additionally, we will undertake the first analysis of the evolution of magnetic fields in a host of objects and determine their connection to other physical parameters. We will elucidate properties of the medium surrounding AGN potentially to redshift 10 and unveil the first polarisation spectral energy distributions. Furthermore, the phase 1 SKA will place tight constraints on the magneto-ionic turbulence in our Galaxy, nearby galaxies and perhaps even make the first detection of the magnetic component of the cosmic web itself.

In conclusion, the SKA's sensitivity, bandwidths and field of view will provide a transformational capability with which to explore magnetic fields. The immensely rich polarimetric data provided by the SKA will allow us to finally address fundamental questions relating to the original, evolution and influence of cosmic magnetic fields on all scales in the Universe, and will mark the start of the \emph{"Era of Precision Magnetism Science"}.

\setlength{\bibsep}{0.0pt}
\bibliographystyle{apj}
\bibliography{BB}

\begin{thebibliography}{}
\expandafter\ifx\csname natexlab\endcsname\relax\def\natexlab#1{#1}\fi

\bibitem[{{Agudo} {et~al.}(2015){Agudo}, {Boettcher}, {Falcke},
  {Georganopoulos}, {Ghisellini}, {Giovannini}, {Giroletti}, {Gomez},
  {Gurvits}, {Laing}, {Lister}, {Marti}, {Meyer}, {Mizuno}, {O'Sullivan},
  {Padovani}, {Paragi}, {Perucho}, {Schleicher}, {Stawarz}, {Vlahakis}, \&
  {Wardle}}]{Agudo15}
{Agudo}, I., {Boettcher}, M., {Falcke}, H., {et~al.} 2015, `Studies of
  Relativistic Jets in Active Galactic Nuclei with SKA' in proc. {\em Advancing
  Astrophysics with the Square Kilometre Array}, PoS(AASKA14)093,
  arXiv:1501.00420

\bibitem[{{Akahori} {et~al.}(2014{\natexlab{a}}){Akahori}, {Gaensler}, \&
  {Ryu}}]{Akahori14a}
{Akahori}, T., {Gaensler}, B.~M., \& {Ryu}, D. 2014{\natexlab{a}}, \apj, 790,
  123

\bibitem[{{Akahori} {et~al.}(2014{\natexlab{b}}){Akahori}, {Kumazaki},
  {Takahashi}, \& {Ryu}}]{Akahori14b}
{Akahori}, T., {Kumazaki}, K., {Takahashi}, K., \& {Ryu}, D.
  2014{\natexlab{b}}, Publications of the Astronomical Society of Japan, 66, 65

\bibitem[{{Beck}(2007)}]{Beck07}
{Beck}, R. 2007, \aap, 470, 539

\bibitem[{{Beck}(2015{\natexlab{a}})}]{Beck2015c}
{Beck}, R. 2015{\natexlab{a}}, in Astrophysics and Space Science Library, Vol.
  407, Astrophysics and Space Science Library, ed. A.~{Lazarian}, E.~M. {de
  Gouveia Dal Pino}, \& C.~{Melioli}, 507

\bibitem[{{Beck}(2015{\natexlab{b}})}]{Beck15b}
---. 2015{\natexlab{b}}, ArXiv e-prints, astro-ph/1502.05439, arXiv:1502.05439

\bibitem[{{Beck} \& {Hoernes}(1996)}]{Beck96}
{Beck}, R., \& {Hoernes}, P. 1996, Nature, 379, 47

\bibitem[{{Beck} {et~al.}(2015){Beck}, {Bomans}, {Colafrancesco}, {Dettmar},
  {Ferri{\`e}re}, {Fletcher}, {Heald}, {Heesen}, {Horellou}, {Krause}, {Lou},
  {Mao}, {Paladino}, {Schinnerer}, {Sokoloff}, {Stil}, \&
  {Tabatabaei}}]{Beck15}
{Beck}, R., {Bomans}, D., {Colafrancesco}, S., {et~al.} 2015, `Structure,
  dynamical impact and origin of magnetic fields in nearby galaxies in the SKA
  era' in proc. {\em Advancing Astrophysics with the Square Kilometre Array},
  PoS(AASKA14)094, arXiv:1501.00385

\bibitem[{{Bernardi} {et~al.}(2013){Bernardi}, {Greenhill}, {Mitchell}, {Ord},
  {Hazelton}, {Gaensler}, {de Oliveira-Costa}, {Morales}, {Udaya Shankar},
  {Subrahmanyan}, {Wayth}, {Lenc}, {Williams}, {Arcus}, {Arora}, {Barnes},
  {Bowman}, {Briggs}, {Bunton}, {Cappallo}, {Corey}, {Deshpande}, {deSouza},
  {Emrich}, {Goeke}, {Herne}, {Hewitt}, {Johnston-Hollitt}, {Kaplan}, {Kasper},
  {Kincaid}, {Koenig}, {Kratzenberg}, {Lonsdale}, {Lynch}, {McWhirter},
  {Morgan}, {Oberoi}, {Pathikulangara}, {Prabu}, {Remillard}, {Rogers},
  {Roshi}, {Salah}, {Sault}, {Srivani}, {Stevens}, {Tingay}, {Waterson},
  {Webster}, {Whitney}, {Williams}, \& {Wyithe}}]{Bernardi13}
{Bernardi}, G., {Greenhill}, L.~J., {Mitchell}, D.~A., {et~al.} 2013, \apj,
  771, 105

\bibitem[{{Bonafede} {et~al.}(2010){Bonafede}, {Feretti}, {Murgia}, {Govoni},
  {Giovannini}, {Dallacasa}, {Dolag}, \& {Taylor}}]{Bonafede10}
{Bonafede}, A., {Feretti}, L., {Murgia}, M., {et~al.} 2010, \aap, 513, A30

\bibitem[{{Bonafede} {et~al.}(2013){Bonafede}, {Vazza}, {Br{\"u}ggen},
  {Murgia}, {Govoni}, {Feretti}, {Giovannini}, \& {Ogrean}}]{Bonafede13}
{Bonafede}, A., {Vazza}, F., {Br{\"u}ggen}, M., {et~al.} 2013, \mnras, 433,
  3208

\bibitem[{{Bonafede} {et~al.}(2009){Bonafede}, {Feretti}, {Giovannini},
  {Govoni}, {Murgia}, {Taylor}, {Ebeling}, {Allen}, {Gentile}, \&
  {Pihlstr{\"o}m}}]{Bonafede09}
{Bonafede}, A., {Feretti}, L., {Giovannini}, G., {et~al.} 2009, \aap, 503, 707

\bibitem[{{Bonafede} {et~al.}(2015){Bonafede}, {Vazza}, {Br{\"u}ggen},
  {Akahori}, {Carretti}, {Colafrancesco}, {Feretti}, {Ferrari}, {Giovannini},
  {Govoni}, {Johnston-Hollitt}, {Murgia}, {Rudnick}, {Scaife}, \&
  {Vacca}}]{Bonafede15}
{Bonafede}, A., {Vazza}, F., {Br{\"u}ggen}, M., {et~al.} 2015, `Unravelling the
  origin of large-scale magnetic fields in galaxy clusters and beyond through
  Faraday Rotation Measures with the SKA' in proc. {\em Advancing Astrophysics
  with the Square Kilometre Array}, PoS(AASKA14)095, arXiv:1501.00321

\bibitem[{{Carretti} {et~al.}(2013){Carretti}, {Crocker}, {Staveley-Smith},
  {Haverkorn}, {Purcell}, {Gaensler}, {Bernardi}, {Kesteven}, \&
  {Poppi}}]{Carretti13}
{Carretti}, E., {Crocker}, R.~M., {Staveley-Smith}, L., {et~al.} 2013, Nature,
  493, 66

\bibitem[{{Cassano} {et~al.}(2015){Cassano}, {Bernardi}, {Brunetti},
  {Br{\"u}ggen}, {Clarke}, {Dallacasa}, {Dolag}, {Ettori}, {Giacintucci},
  {Giocoli}, {Gitti}, {Johnston-Hollitt}, {Kale}, {Markevitch}, {Norris},
  {Pandey-Pommier}, {Pratt}, {R{\"o}ttgering}, \& {Venturi}}]{Cassano15}
{Cassano}, R., {Bernardi}, G., {Brunetti}, G., {et~al.} 2015, `Cluster Radio
  Halos at the crossroads between astrophysics and cosmology in the SKA era' in
  proc. {\em Advancing Astrophysics with the Square Kilometre Array},
  PoS(AASKA14)073, arXiv:1412.5940

\bibitem[{{Chamandy} {et~al.}(2015){Chamandy}, {Shukurov}, \&
  {Subramanian}}]{Chamandy15}
{Chamandy}, L., {Shukurov}, A., \& {Subramanian}, K. 2015, \mnras, 446, L6

\bibitem[{{Clarke} {et~al.}(2001){Clarke}, {Kronberg}, \&
  {B{\"o}hringer}}]{Clarke01}
{Clarke}, T.~E., {Kronberg}, P.~P., \& {B{\"o}hringer}, H. 2001, \apjl, 547,
  L111

\bibitem[{{Colafrancesco} {et~al.}(2015){Colafrancesco}, {Regis},
  {Marchegiani}, {Beck}, {Beck}, {Zechlin}, {Lobanov}, \&
  {Horns}}]{Colafrancesco15}
{Colafrancesco}, S., {Regis}, M., {Marchegiani}, P., {et~al.} 2015, `Probing
  the nature of Dark Matter with the SKA' in proc. {\em Advancing Astrophysics
  with the Square Kilometre Array}, PoS(AASKA14)100, arXiv:1502.03738

\bibitem[{{Dickinson} {et~al.}(2015){Dickinson}, {Beck}, {Crocker}, {Crutcher},
  {Davies}, {Ferriere}, {Fuller}, {Jaffe}, {Jones}, {Leahy}, {Murphy}, {Peel},
  {Orlando}, {Porter}, {Protheroe}, {Robishaw}, {Strong}, {Watson}, \&
  {Yusef-Zadeh}}]{Dickinson15}
{Dickinson}, C., {Beck}, R., {Crocker}, R., {et~al.} 2015, `SKA studies of
  in-situ synchrotron radiation from molecular clouds' in proc. {\em Advancing
  Astrophysics with the Square Kilometre Array}, PoS(AASKA14)102,
  arXiv:1501.00804

\bibitem[{{Feain} {et~al.}(2009){Feain}, {Ekers}, {Murphy}, {Gaensler},
  {Macquart}, {Norris}, {Cornwell}, {Johnston-Hollitt}, {Ott}, \&
  {Middelberg}}]{Feain09}
{Feain}, I.~J., {Ekers}, R.~D., {Murphy}, T., {et~al.} 2009, \apj, 707, 114

\bibitem[{{Feretti} \& {Johnston-Hollitt}(2004)}]{Feretti04}
{Feretti}, L., \& {Johnston-Hollitt}, M. 2004, \nar, 48, 1145

\bibitem[{{Fomalont} {et~al.}(1989){Fomalont}, {Ebneter}, {van Breugel}, \&
  {Ekers}}]{Fomalont89}
{Fomalont}, E.~B., {Ebneter}, K.~A., {van Breugel}, W.~J.~M., \& {Ekers}, R.~D.
  1989, \apjl, 346, L17

\bibitem[{{Gaensler} {et~al.}(2004){Gaensler}, {Beck}, \&
  {Feretti}}]{Gaensler04}
{Gaensler}, B.~M., {Beck}, R., \& {Feretti}, L. 2004, \nar, 48, 1003

\bibitem[{{Gaensler} {et~al.}(2005){Gaensler}, {Haverkorn}, {Staveley-Smith},
  {Dickey}, {McClure-Griffiths}, {Dickel}, \& {Wolleben}}]{Gaensler05}
{Gaensler}, B.~M., {Haverkorn}, M., {Staveley-Smith}, L., {et~al.} 2005,
  Science, 307, 1610

\bibitem[{{Gaensler} {et~al.}(2010){Gaensler}, {Landecker}, {Taylor}, \&
  {POSSUM Collaboration}}]{Gaensler10}
{Gaensler}, B.~M., {Landecker}, T.~L., {Taylor}, A.~R., \& {POSSUM
  Collaboration}. 2010, in Bulletin of the American Astronomical Society,
  Vol.~42, American Astronomical Society Meeting Abstracts no. 215, 470.13

\bibitem[{{Gaensler} {et~al.}(2011){Gaensler}, {Haverkorn}, {Burkhart},
  {Newton-McGee}, {Ekers}, {Lazarian}, {McClure-Griffiths}, {Robishaw},
  {Dickey}, \& {Green}}]{Gaensler11}
{Gaensler}, B.~M., {Haverkorn}, M., {Burkhart}, B., {et~al.} 2011, Nature, 478,
  214

\bibitem[{{Gaensler} {et~al.}(2015){Gaensler}, {Agudo}, {Akahori}, {Banfield},
  {Beck}, {Carretti}, {Farnes}, {Haverkorn}, {Heald}, {Jones}, {Landecker},
  {Mao}, {Norris}, {O'Sullivan}, {Rudnick}, {Schnitzeler}, {Seymour}, \&
  {Sun}}]{Gaensler15}
{Gaensler}, B.~M., {Agudo}, I., {Akahori}, T., {et~al.} 2015, `Broadband
  Polarimetry with the Square Kilometre Array: A Unique Astrophysical Probe' in
  proc. {\em Advancing Astrophysics with the Square Kilometre Array},
  PoS(AASKA14)103, arXiv:1501.00626

\bibitem[{{Gent} {et~al.}(2013){Gent}, {Shukurov}, {Sarson}, {Fletcher}, \&
  {Mantere}}]{Gent13}
{Gent}, F.~A., {Shukurov}, A., {Sarson}, G.~R., {Fletcher}, A., \& {Mantere},
  M.~J. 2013, \mnras, 430, L40

\bibitem[{{Giovannini} {et~al.}(2015){Giovannini}, {Bonafede}, {Brown},
  {Feretti}, {Ferrari}, {Gitti}, {Govoni}, {Murgia}, \& {Vacca}}]{Giovannini15}
{Giovannini}, G., {Bonafede}, A., {Brown}, S., {et~al.} 2015, `Mega-parsec
  scale magnetic fields in low density regions in the SKA era: filaments
  connecting galaxy clusters and groups' in proc. {\em Advancing Astrophysics
  with the Square Kilometre Array}, PoS(AASKA14)104, arXiv:1501.01023

\bibitem[{{Govoni} {et~al.}(2005){Govoni}, {Murgia}, {Feretti}, {Giovannini},
  {Dallacasa}, \& {Taylor}}]{Govoni05}
{Govoni}, F., {Murgia}, M., {Feretti}, L., {et~al.} 2005, \aap, 430, L5

\bibitem[{{Govoni} {et~al.}(2013){Govoni}, {Murgia}, {Xu}, {Li}, {Norman},
  {Feretti}, {Giovannini}, \& {Vacca}}]{Govoni13}
{Govoni}, F., {Murgia}, M., {Xu}, H., {et~al.} 2013, \aap, 554, A102

\bibitem[{{Govoni} {et~al.}(2015){Govoni}, {Murgia}, {Xu}, {Li}, {Norman},
  {Feretti}, {Giovannini}, {Vacca}, {Bernardi}, {Bonafede}, {Brunetti},
  {Carretti}, {Colafrancesco}, {Donnert}, {Ferrari}, {Gitti}, {Iapichino},
  {Johnston-Hollitt}, {Pizzo}, \& {Rudnick}}]{Govoni15}
---. 2015, `Cluster magnetic fields through the study of polarized radio halos
  in the SKA era' in proc. {\em Advancing Astrophysics with the Square
  Kilometre Array}, PoS(AASKA14)105, arXiv:1501.00389

\bibitem[{{Guidetti} {et~al.}(2012){Guidetti}, {Laing}, {Croston}, {Bridle}, \&
  {Parma}}]{Guidetti12}
{Guidetti}, D., {Laing}, R.~A., {Croston}, J.~H., {Bridle}, A.~H., \& {Parma},
  P. 2012, \mnras, 423, 1335

\bibitem[{{Guidetti} {et~al.}(2008){Guidetti}, {Murgia}, {Govoni},
  {et~al.}}]{Guidetti08}
{Guidetti}, D., {Murgia}, M., {Govoni}, F., {et~al.} 2008, \aap, 483, 699

\bibitem[{{Hales} {et~al.}(2014){Hales}, {Norris}, {Gaensler}, {Middelberg},
  {Chow}, {Hopkins}, {Huynh}, {Lenc}, \& {Mao}}]{Hales14}
{Hales}, C.~A., {Norris}, R.~P., {Gaensler}, B.~M., {et~al.} 2014, \mnras, 441,
  2555

\bibitem[{{Han} {et~al.}(2015){Han}, {van Straten}, {Lazio}, {Deller}, {Sobey},
  {Xu}, {Schnitzeler}, {Imai}, {Chatterjee}, {Macquart}, {Kramer}, \&
  {Cordes}}]{Han15}
{Han}, J.~L., {van Straten}, W., {Lazio}, T.~J.~W., {et~al.} 2015,
  `Three-dimensional Tomography of the Galactic and Extragalactic Magnetoionic
  Medium with the SKA' in proc. {\em Advancing Astrophysics with the Square
  Kilometre Array}, PoS(AASKA14)041, arXiv:1412.8749

\bibitem[{{Harvey-Smith} {et~al.}(2011){Harvey-Smith}, {Madsen}, \&
  {Gaensler}}]{LHS11}
{Harvey-Smith}, L., {Madsen}, G.~J., \& {Gaensler}, B.~M. 2011, \apj, 736, 83

\bibitem[{{Haverkorn} {et~al.}(2006){Haverkorn}, {Gaensler},
  {McClure-Griffiths}, {Dickey}, \& {Green}}]{Haverkorn06}
{Haverkorn}, M., {Gaensler}, B.~M., {McClure-Griffiths}, N.~M., {Dickey},
  J.~M., \& {Green}, A.~J. 2006, \apjs, 167, 230

\bibitem[{{Haverkorn} {et~al.}(2015){Haverkorn}, {Akahori}, {Carretti},
  {Ferriere}, {Frick}, {Gaensler}, {Heald}, {Johnston-Hollitt}, {Jones},
  {Landecker}, {Mao}, {Noutsos}, {Oppermann}, {Reich}, {Robishaw}, {Scaife},
  {Schnitzeler}, {Stepanov}, {Sun}, {Taylor}, \& {for the SKA Cosmic Magnetism
  Working Group}}]{Haverkorn15}
{Haverkorn}, M., {Akahori}, T., {Carretti}, E., {et~al.} 2015, `Measuring
  magnetism in the Milky Way with the Square Kilometre Array' in proc. {\em
  Advancing Astrophysics with the Square Kilometre Array}, PoS(AASKA14)096,
  arXiv:1501.00416

\bibitem[{{Heald} {et~al.}(2015){Heald}, {Beck}, {de Blok}, {Dettmar},
  {Fletcher}, {Gaensler}, {Haverkorn}, {Heesen}, {Horellou}, {Krause}, {Mao},
  {Oppermann}, {Scaife}, {Sokoloff}, {Stil}, {Tabatabaei}, {Takahashi},
  {Taylor}, \& {Williams}}]{Heald15}
{Heald}, G., {Beck}, R., {de Blok}, W.~J.~G., {et~al.} 2015, `Magnetic Field
  Tomography in Nearby Galaxies with the Square Kilometre Array' in proc. {\em
  Advancing Astrophysics with the Square Kilometre Array}, PoS(AASKA14)106,
  arXiv:1501.00408

\bibitem[{{Heald}(2012)}]{Heald12}
{Heald}, G.~H. 2012, \apjl, 754, L35

\bibitem[{{Hennessy} {et~al.}(1989){Hennessy}, {Owen}, \& {Eilek}}]{Hennessy89}
{Hennessy}, G.~S., {Owen}, F.~N., \& {Eilek}, J.~A. 1989, \apj, 347, 144

\bibitem[{{Hill} {et~al.}(2008){Hill}, {Benjamin}, {Kowal}, {Reynolds},
  {Haffner}, \& {Lazarian}}]{Hill08}
{Hill}, A.~S., {Benjamin}, R.~A., {Kowal}, G., {et~al.} 2008, \apj, 686, 363

\bibitem[{{Hotan} {et~al.}(2014){Hotan}, {Bunton}, {Harvey-Smith}, {Humphreys},
  {Jeffs}, {Shimwell}, {Tuthill}, {Voronkov}, {Allen}, {Amy}, {Ardern},
  {Axtens}, {Ball}, {Bannister}, {Barker}, {Bateman}, {Beresford}, {Bock},
  {Bolton}, {Bowen}, {Boyle}, {Braun}, {Broadhurst}, {Brodrick}, {Brooks},
  {Brothers}, {Brown}, {Cantrall}, {Carrad}, {Chapman}, {Cheng}, {Chippendale},
  {Chung}, {Cooray}, {Cornwell}, {Davis}, {de Souza}, {DeBoer}, {Diamond},
  {Edwards}, {Ekers}, {Feain}, {Ferris}, {Forsyth}, {Gough}, {Grancea},
  {Gupta}, {Guzman}, {Hampson}, {Haskins}, {Hay}, {Hayman}, {Hoyle}, {Jacka},
  {Jackson}, {Jackson}, {Jeganathan}, {Johnston}, {Joseph}, {Kendall},
  {Kesteven}, {Kiraly}, {Koribalski}, {Leach}, {Lenc}, {Lensson}, {Li},
  {Mackay}, {Macleod}, {Maher}, {Marquarding}, {McClure-Griffiths},
  {McConnell}, {Mickle}, {Mirtschin}, {Norris}, {Neuhold}, {Ng}, {O'Sullivan},
  {Pathikulangara}, {Pearce}, {Phillips}, {Qiao}, {Reynolds}, {Rispler},
  {Roberts}, {Roxby}, {Schinckel}, {Shaw}, {Shields}, {Storey}, {Sweetnam},
  {Troup}, {Turner}, {Tzioumis}, {Westmeier}, {Whiting}, {Wilson}, {Wilson},
  {Wormnes}, \& {Wu}}]{Hotan14}
{Hotan}, A.~W., {Bunton}, J.~D., {Harvey-Smith}, L., {et~al.} 2014,
  Publications of the Astronomical Society of Australia, 31, 41

\bibitem[{{Janssen} {et~al.}(2015){Janssen}, {Hobbs}, {McLaughlin}, {Bassa},
  {Deller}, {Kramer}, {Lee}, {Mingarelli}, {Rosado}, {Sanidas}, {Sesana},
  {Shao}, {Stairs}, {Stappers}, \& {Verbiest}}]{Janssen15}
{Janssen}, G.~H., {Hobbs}, G., {McLaughlin}, M., {et~al.} 2015, `Stacking for
  Cosmic Magnetism with SKA Surveys' in proc. {\em Advancing Astrophysics with
  the Square Kilometre Array}, PoS(AASKA14)037, arXiv:1501.00127

\bibitem[{{Jeli{\'c}} {et~al.}(2014){Jeli{\'c}}, {de Bruyn}, {Mevius},
  {Abdalla}, {Asad}, {Bernardi}, {Brentjens}, {Bus}, {Chapman}, {Ciardi},
  {Daiboo}, {Fernandez}, {Ghosh}, {Harker}, {Jensen}, {Kazemi}, {Koopmans},
  {Labropoulos}, {Martinez-Rubi}, {Mellema}, {Offringa}, {Pandey}, {Patil},
  {Thomas}, {Vedantham}, {Veligatla}, {Yatawatta}, {Zaroubi}, {Alexov},
  {Anderson}, {Avruch}, {Beck}, {Bell}, {Bentum}, {Best}, {Bonafede},
  {Bregman}, {Breitling}, {Broderick}, {Brouw}, {Br{\"u}ggen}, {Butcher},
  {Conway}, {de Gasperin}, {de Geus}, {Deller}, {Dettmar}, {Duscha},
  {Eisl{\"o}ffel}, {Engels}, {Falcke}, {Fallows}, {Fender}, {Ferrari},
  {Frieswijk}, {Garrett}, {Grie{\ss}meier}, {Gunst}, {Hamaker}, {Hassall},
  {Haverkorn}, {Heald}, {Hessels}, {Hoeft}, {H{\"o}randel}, {Horneffer}, {van
  der Horst}, {Iacobelli}, {Juette}, {Karastergiou}, {Kondratiev}, {Kramer},
  {Kuniyoshi}, {Kuper}, {van Leeuwen}, {Maat}, {Mann}, {McKay-Bukowski},
  {McKean}, {Munk}, {Nelles}, {Norden}, {Paas}, {Pandey-Pommier}, {Pietka},
  {Pizzo}, {Polatidis}, {Reich}, {R{\"o}ttgering}, {Rowlinson}, {Scaife},
  {Schwarz}, {Serylak}, {Smirnov}, {Steinmetz}, {Stewart}, {Tagger}, {Tang},
  {Tasse}, {ter Veen}, {Thoudam}, {Toribio}, {Vermeulen}, {Vocks}, {van
  Weeren}, {Wijers}, {Wijnholds}, {Wucknitz}, \& {Zarka}}]{Jelic14}
{Jeli{\'c}}, V., {de Bruyn}, A.~G., {Mevius}, M., {et~al.} 2014, \aap, 568,
  A101

\bibitem[{{Johnston-Hollitt}(2003)}]{MJH03}
{Johnston-Hollitt}, M. 2003, PhD thesis, University of Adelaide

\bibitem[{{Johnston-Hollitt}(2004)}]{MJH04a}
{Johnston-Hollitt}, M. 2004, in The Riddle of Cooling Flows in Galaxies and
  Clusters of galaxies, ed. T.~{Reiprich}, J.~{Kempner}, \& N.~{Soker}, 51

\bibitem[{{Johnston-Hollitt} {et~al.}(2015{\natexlab{a}}){Johnston-Hollitt},
  {Dehghan}, \& {Pratley}}]{MJH15a}
{Johnston-Hollitt}, M., {Dehghan}, S., \& {Pratley}, L. 2015{\natexlab{a}},
  `Using Tailed Radio Galaxies to Probe the Environment and Magnetic Field of
  Galaxy Clusters in the SKA Era' in proc. {\em Advancing Astrophysics with the
  Square Kilometre Array}, PoS(AASKA14)101, arXiv:1501.00761

\bibitem[{{Johnston-Hollitt} {et~al.}(2015{\natexlab{b}}){Johnston-Hollitt},
  {Dehghan}, \& {Pratley}}]{MJH15b}
{Johnston-Hollitt}, M., {Dehghan}, S., \& {Pratley}, L. 2015{\natexlab{b}}, in
  IAU Symposium, Vol. 313, IAU Symposium, ed. F.~{Massaro}, C.~C. {Cheung},
  E.~{Lopez}, \& A.~{Siemiginowska}, 321--326

\bibitem[{{Johnston-Hollitt} \& {Ekers}(2004)}]{MJHetal04}
{Johnston-Hollitt}, M., \& {Ekers}, R.~D. 2004, ArXiv Astrophysics e-prints,
  astro-ph/0411045, astro-ph/0411045

\bibitem[{{Johnston-Hollitt} {et~al.}(2004){Johnston-Hollitt}, {Hollitt}, \&
  {Ekers}}]{MJH04b}
{Johnston-Hollitt}, M., {Hollitt}, C.~P., \& {Ekers}, R.~D. 2004, in The
  Magnetized Interstellar Medium, ed. B.~{Uyaniker}, W.~{Reich}, \&
  R.~{Wielebinski}, 13--18

\bibitem[{{Keane} {et~al.}(2015){Keane}, {Bhattacharyya}, {Kramer}, {Stappers},
  {Bates}, {Burgay}, {Chatterjee}, {Champion}, {Eatough}, {Hessels}, {Janssen},
  {Lee}, {van Leeuwen}, {Margueron}, {Oertel}, {Possenti}, {Ransom},
  {Theureau}, \& {Torne}}]{Keane15}
{Keane}, E.~F., {Bhattacharyya}, B., {Kramer}, M., {et~al.} 2015, `Giant radio
  galaxies as probes of the ambient WHIM in the era of the SKA' in proc. {\em
  Advancing Astrophysics with the Square Kilometre Array}, PoS(AASKA14)040,
  arXiv:1501.00056

\bibitem[{{Kim} {et~al.}(1991){Kim}, {Tribble}, \& {Kronberg}}]{Kim91}
{Kim}, K.-T., {Tribble}, P.~C., \& {Kronberg}, P.~P. 1991, \apj, 379, 80

\bibitem[{{Krause}(2009)}]{Krause09}
{Krause}, M. 2009, in Revista Mexicana de Astronomia y Astrofisica, vol. 27,
  Vol.~36, Revista Mexicana de Astronomia y Astrofisica Conference Series,
  25--29

\bibitem[{{Laing}(2015)}]{Laing15}
{Laing}, R.~A. 2015, `Kinematics and Dynamics of kiloparsec-scale Jets in Radio
  Galaxies with SKA' in proc. {\em Advancing Astrophysics with the Square
  Kilometre Array}, PoS(AASKA14)107, arXiv:1501.00452

\bibitem[{{Lou} \& {Fan}(1998)}]{Lou98}
{Lou}, Y.-Q., \& {Fan}, Z. 1998, \apj, 493, 102

\bibitem[{{Macquart} {et~al.}(2015){Macquart}, {Keane}, {Grainge}, {McQuinn},
  {Fender}, {Hessels}, {Deller}, {Bhat}, {Breton}, {Chatterjee}, {Law},
  {Lorimer}, {Ofek}, {Pietka}, {Spitler}, {Stappers}, \& {Trott}}]{Macquart15}
{Macquart}, J.-P., {Keane}, E., {Grainge}, K., {et~al.} 2015, `Fast Transients
  at Cosmological Distances with the SKA' in proc. {\em Advancing Astrophysics
  with the Square Kilometre Array}, PoS(AASKA14)055, arXiv:1501.07535

\bibitem[{{McClure-Griffiths} {et~al.}(2010){McClure-Griffiths}, {Madsen},
  {Gaensler}, {McConnell}, \& {Schnitzeler}}]{McG10}
{McClure-Griffiths}, N.~M., {Madsen}, G.~J., {Gaensler}, B.~M., {McConnell},
  D., \& {Schnitzeler}, D.~H.~F.~M. 2010, \apj, 725, 275

\bibitem[{{Moss} {et~al.}(2015){Moss}, {Stepanov}, {Krause}, {Beck}, \&
  {Sokoloff}}]{Moss15}
{Moss}, D., {Stepanov}, R., {Krause}, M., {Beck}, R., \& {Sokoloff}, D. 2015,
  ArXiv e-prints, astro-ph/1504.07824, arXiv:1504.07824

\bibitem[{{Oppermann} {et~al.}(2012){Oppermann}, {Junklewitz}, {Robbers},
  {Bell}, {En{\ss}lin}, {Bonafede}, {Braun}, {Brown}, {Clarke}, {Feain},
  {Gaensler}, {Hammond}, {Harvey-Smith}, {Heald}, {Johnston-Hollitt}, {Klein},
  {Kronberg}, {Mao}, {McClure-Griffiths}, {O'Sullivan}, {Pratley}, {Robishaw},
  {Roy}, {Schnitzeler}, {Sotomayor-Beltran}, {Stevens}, {Stil}, {Sunstrum},
  {Tanna}, {Taylor}, \& {Van Eck}}]{Oppermann12}
{Oppermann}, N., {Junklewitz}, H., {Robbers}, G., {et~al.} 2012, \aap, 542, A93

\bibitem[{{Oppermann} {et~al.}(2015){Oppermann}, {Junklewitz}, {Greiner},
  {En{\ss}lin}, {Akahori}, {Carretti}, {Gaensler}, {Goobar}, {Harvey-Smith},
  {Johnston-Hollitt}, {Pratley}, {Schnitzeler}, {Stil}, \&
  {Vacca}}]{Oppermann15}
{Oppermann}, N., {Junklewitz}, H., {Greiner}, M., {et~al.} 2015, \aap, 575,
  A118

\bibitem[{{O'Sullivan} {et~al.}(2013){O'Sullivan}, {Feain},
  {McClure-Griffiths}, {Ekers}, {Carretti}, {Robishaw}, {Mao}, {Gaensler},
  {Bland-Hawthorn}, \& {Stawarz}}]{O'Sullivan13}
{O'Sullivan}, S.~P., {Feain}, I.~J., {McClure-Griffiths}, N.~M., {et~al.} 2013,
  \apj, 764, 162

\bibitem[{{Peng} {et~al.}(2015){Peng}, {Chen}, \& {Strom}}]{Peng15}
{Peng}, B., {Chen}, R.-R., \& {Strom}, R. 2015, `Giant radio galaxies as probes
  of the ambient WHIM in the era of the SKA' in proc. {\em Advancing
  Astrophysics with the Square Kilometre Array}, PoS(AASKA14)109,
  arXiv:1501.00407

\bibitem[{{Pizzo} {et~al.}(2011){Pizzo}, {de Bruyn}, {Bernardi}, \&
  {Brentjens}}]{Pizzo11}
{Pizzo}, R.~F., {de Bruyn}, A.~G., {Bernardi}, G., \& {Brentjens}, M.~A. 2011,
  \aap, 525, A104

\bibitem[{{Prandoni} \& {Seymour}(2015)}]{Prandoni15}
{Prandoni}, I., \& {Seymour}, N. 2015, `Revealing the Physics and Evolution of
  Galaxies and Galaxy Clusters with SKA Continuum Surveys' in proc. {\em
  Advancing Astrophysics with the Square Kilometre Array}, PoS(AASKA14)067,
  arXiv:1412.6512

\bibitem[{{Pratley} {et~al.}(2013){Pratley}, {Johnston-Hollitt}, {Dehghan}, \&
  {Sun}}]{Pratley13}
{Pratley}, L., {Johnston-Hollitt}, M., {Dehghan}, S., \& {Sun}, M. 2013,
  \mnras, 432, 243

\bibitem[{{Purcell} {et~al.}(2015){Purcell}, {Gaensler}, {Sun}, {Carretti},
  {Bernardi}, {Haverkorn}, {Kesteven}, {Poppi}, {Schnitzeler}, \&
  {Staveley-Smith}}]{Purcell15}
{Purcell}, C.~R., {Gaensler}, B.~M., {Sun}, X.~H., {et~al.} 2015, \apj, 804, 22

\bibitem[{{Reich} {et~al.}(2004){Reich}, {F{\"u}rst}, {Reich}, {Uyaniker},
  {Wielebinski}, \& {Wolleben}}]{Reich04}
{Reich}, W., {F{\"u}rst}, E., {Reich}, P., {et~al.} 2004, in The Magnetized
  Interstellar Medium, ed. B.~{Uyaniker}, W.~{Reich}, \& R.~{Wielebinski},
  45--50

\bibitem[{{Robishaw} {et~al.}(2015){Robishaw}, {Green}, {Surcis}, {Vlemmings},
  {Richards}, {Etoka}, {Bourke}, {Fish}, {Gray}, {Imai}, {Kramer}, {McBride},
  {Momjian}, {Sarma}, \& {Zijlstra}}]{Robishaw15}
{Robishaw}, T., {Green}, J.~A., {Surcis}, G., {et~al.} 2015, `Measuring
  Magnetic Fields Near and Far with the SKA via the Zeeman Effect' in proc.
  {\em Advancing Astrophysics with the Square Kilometre Array},
  PoS(AASKA14)110, arXiv:1503.01779

\bibitem[{{Rudnick} \& {Owen}(2014)}]{Rudnick14}
{Rudnick}, L., \& {Owen}, F.~N. 2014, \apj, 785, 45

\bibitem[{{Ryu} {et~al.}(2012){Ryu}, {Schleicher}, {Treumann}, {Tsagas}, \&
  {Widrow}}]{Ryu12}
{Ryu}, D., {Schleicher}, D.~R.~G., {Treumann}, R.~A., {Tsagas}, C.~G., \&
  {Widrow}, L.~M. 2012, Space Science Reviews, 166, 1

\bibitem[{{Schnitzeler}(2010)}]{Schnitzeler10}
{Schnitzeler}, D.~H.~F.~M. 2010, \mnras, 409, L99

\bibitem[{{Smits} {et~al.}(2011){Smits}, {Tingay}, {Wex}, {Kramer}, \&
  {Stappers}}]{Smits11}
{Smits}, R., {Tingay}, S.~J., {Wex}, N., {Kramer}, M., \& {Stappers}, B. 2011,
  \aap, 528, A108

\bibitem[{{Stil} \& {Keller}(2015)}]{Stil15}
{Stil}, J.~M., \& {Keller}, B.~W. 2015, `Stacking for Cosmic Magnetism with SKA
  Surveys' in proc. {\em Advancing Astrophysics with the Square Kilometre
  Array}, PoS(AASKA14)112, arXiv:1501.00390

\bibitem[{{Stil} {et~al.}(2014){Stil}, {Keller}, {George}, \&
  {Taylor}}]{Stil14}
{Stil}, J.~M., {Keller}, B.~W., {George}, S.~J., \& {Taylor}, A.~R. 2014, \apj,
  787, 99

\bibitem[{{Stil} {et~al.}(2011){Stil}, {Taylor}, \& {Sunstrum}}]{Stil11}
{Stil}, J.~M., {Taylor}, A.~R., \& {Sunstrum}, C. 2011, \apj, 726, 4

\bibitem[{{Stutz} {et~al.}(2014){Stutz}, {Rosolowsky}, {Kothes}, \&
  {Landecker}}]{Stutz14}
{Stutz}, R.~A., {Rosolowsky}, E.~W., {Kothes}, R., \& {Landecker}, T.~L. 2014,
  \apj, 787, 34

\bibitem[{{Tabatabaei} {et~al.}(2013){Tabatabaei}, {Schinnerer}, {Murphy},
  {Beck}, {Groves}, {Meidt}, {Krause}, {Rix}, {Sandstrom}, {Crocker},
  {Galametz}, {Helou}, {Wilson}, {Kennicutt}, {Calzetti}, {Draine}, {Aniano},
  {Dale}, {Dumas}, {Engelbracht}, {Gordon}, {Hinz}, {Kreckel}, {Montiel}, \&
  {Roussel}}]{Tabatabaei13}
{Tabatabaei}, F.~S., {Schinnerer}, E., {Murphy}, E.~J., {et~al.} 2013, \aap,
  552, A19

\bibitem[{{Taylor} {et~al.}(2009){Taylor}, {Stil}, \& {Sunstrum}}]{Taylor09}
{Taylor}, A.~R., {Stil}, J.~M., \& {Sunstrum}, C. 2009, \apj, 702, 1230

\bibitem[{{Taylor} {et~al.}(2015){Taylor}, {Agudo}, {Akahori}, {Beck},
  {Gaensler}, {Heald}, {Johnston-Hollitt}, {Langer}, {Rudnick}, {Ryu},
  {Scaife}, {Schleicher}, \& {Stil}}]{Taylor15}
{Taylor}, A.~R., {Agudo}, I., {Akahori}, T., {et~al.} 2015, `SKA Deep
  Polarization and Cosmic Magnetism' in proc. {\em Advancing Astrophysics with
  the Square Kilometre Array}, PoS(AASKA14)113, arXiv:1501.02298

\bibitem[{{Tingay} {et~al.}(2013){Tingay}, {Goeke}, {Bowman}, {Emrich}, {Ord},
  {Mitchell}, {Morales}, {Booler}, {Crosse}, {Wayth}, {Lonsdale}, {Tremblay},
  {Pallot}, {Colegate}, {Wicenec}, {Kudryavtseva}, {Arcus}, {Barnes},
  {Bernardi}, {Briggs}, {Burns}, {Bunton}, {Cappallo}, {Corey}, {Deshpande},
  {Desouza}, {Gaensler}, {Greenhill}, {Hall}, {Hazelton}, {Herne}, {Hewitt},
  {Johnston-Hollitt}, {Kaplan}, {Kasper}, {Kincaid}, {Koenig}, {Kratzenberg},
  {Lynch}, {Mckinley}, {Mcwhirter}, {Morgan}, {Oberoi}, {Pathikulangara},
  {Prabu}, {Remillard}, {Rogers}, {Roshi}, {Salah}, {Sault}, {Udaya-Shankar},
  {Schlagenhaufer}, {Srivani}, {Stevens}, {Subrahmanyan}, {Waterson},
  {Webster}, {Whitney}, {Williams}, {Williams}, \& {Wyithe}}]{Tingay13}
{Tingay}, S.~J., {Goeke}, R., {Bowman}, J.~D., {et~al.} 2013, Publications of
  the Astronomical Society of Australia, 30, 7

\bibitem[{{Vacca} {et~al.}(2015){Vacca}, {Oppermann}, {Ensslin}, {Selig},
  {Junklewitz}, {Greiner}, {Jasche}, {Hales}, {Reinecke}, {Carretti},
  {Feretti}, {Ferrari}, {Giovannini}, {Govoni}, {Horellou}, {Ideguchi},
  {Johnston-Hollitt}, {Murgia}, {Paladino}, {Pizzo}, \& {Anna}}]{Vacca15}
{Vacca}, V., {Oppermann}, N., {Ensslin}, T., {et~al.} 2015, `Statistical
  methods for the analysis of rotation measure grids in large scale structures
  in the SKA era' in proc. {\em Advancing Astrophysics with the Square
  Kilometre Array}, PoS(AASKA14)114, arXiv:1501.00415

\bibitem[{{van Haarlem} {et~al.}(2013){van Haarlem}, {Wise}, {Gunst}, {Heald},
  {McKean}, {Hessels}, {de Bruyn}, {Nijboer}, {Swinbank}, {Fallows},
  {Brentjens}, {Nelles}, {Beck}, {Falcke}, {Fender}, {H{\"o}randel},
  {Koopmans}, {Mann}, {Miley}, {R{\"o}ttgering}, {Stappers}, {Wijers},
  {Zaroubi}, {van den Akker}, {Alexov}, {Anderson}, {Anderson}, {van Ardenne},
  {Arts}, {Asgekar}, {Avruch}, {Batejat}, {B{\"a}hren}, {Bell}, {Bell}, {van
  Bemmel}, {Bennema}, {Bentum}, {Bernardi}, {Best}, {B{\^i}rzan}, {Bonafede},
  {Boonstra}, {Braun}, {Bregman}, {Breitling}, {van de Brink}, {Broderick},
  {Broekema}, {Brouw}, {Br{\"u}ggen}, {Butcher}, {van Cappellen}, {Ciardi},
  {Coenen}, {Conway}, {Coolen}, {Corstanje}, {Damstra}, {Davies}, {Deller},
  {Dettmar}, {van Diepen}, {Dijkstra}, {Donker}, {Doorduin}, {Dromer}, {Drost},
  {van Duin}, {Eisl{\"o}ffel}, {van Enst}, {Ferrari}, {Frieswijk}, {Gankema},
  {Garrett}, {de Gasperin}, {Gerbers}, {de Geus}, {Grie{\ss}meier}, {Grit},
  {Gruppen}, {Hamaker}, {Hassall}, {Hoeft}, {Holties}, {Horneffer}, {van der
  Horst}, {van Houwelingen}, {Huijgen}, {Iacobelli}, {Intema}, {Jackson},
  {Jelic}, {de Jong}, {Juette}, {Kant}, {Karastergiou}, {Koers}, {Kollen},
  {Kondratiev}, {Kooistra}, {Koopman}, {Koster}, {Kuniyoshi}, {Kramer},
  {Kuper}, {Lambropoulos}, {Law}, {van Leeuwen}, {Lemaitre}, {Loose}, {Maat},
  {Macario}, {Markoff}, {Masters}, {McFadden}, {McKay-Bukowski}, {Meijering},
  {Meulman}, {Mevius}, {Middelberg}, {Millenaar}, {Miller-Jones}, {Mohan},
  {Mol}, {Morawietz}, {Morganti}, {Mulcahy}, {Mulder}, {Munk}, {Nieuwenhuis},
  {van Nieuwpoort}, {Noordam}, {Norden}, {Noutsos}, {Offringa}, {Olofsson},
  {Omar}, {Orr{\'u}}, {Overeem}, {Paas}, {Pandey-Pommier}, {Pandey}, {Pizzo},
  {Polatidis}, {Rafferty}, {Rawlings}, {Reich}, {de Reijer}, {Reitsma},
  {Renting}, {Riemers}, {Rol}, {Romein}, {Roosjen}, {Ruiter}, {Scaife}, {van
  der Schaaf}, {Scheers}, {Schellart}, {Schoenmakers}, {Schoonderbeek},
  {Serylak}, {Shulevski}, {Sluman}, {Smirnov}, {Sobey}, {Spreeuw}, {Steinmetz},
  {Sterks}, {Stiepel}, {Stuurwold}, {Tagger}, {Tang}, {Tasse}, {Thomas},
  {Thoudam}, {Toribio}, {van der Tol}, {Usov}, {van Veelen}, {van der Veen},
  {ter Veen}, {Verbiest}, {Vermeulen}, {Vermaas}, {Vocks}, {Vogt}, {de Vos},
  {van der Wal}, {van Weeren}, {Weggemans}, {Weltevrede}, {White}, {Wijnholds},
  {Wilhelmsson}, {Wucknitz}, {Yatawatta}, {Zarka}, {Zensus}, \& {van
  Zwieten}}]{vanHaarlem13}
{van Haarlem}, M.~P., {Wise}, M.~W., {Gunst}, A.~W., {et~al.} 2013, \aap, 556,
  A2

\bibitem[{{Vazza} {et~al.}(2015){Vazza}, {Ferrari}, {Bonafede}, {Br{\"u}ggen},
  {Gheller}, {Braun}, \& {Brown}}]{Vazza15}
{Vazza}, F., {Ferrari}, C., {Bonafede}, A., {et~al.} 2015, `Filaments of the
  radio cosmic web: opportunities and challenges for SKA' in proc. {\em
  Advancing Astrophysics with the Square Kilometre Array}, PoS(AASKA14)097,
  arXiv:1501.00315

\bibitem[{{Vikhlinin} {et~al.}(2001){Vikhlinin}, {Markevitch}, \&
  {Murray}}]{Vikhlinin01}
{Vikhlinin}, A., {Markevitch}, M., \& {Murray}, S.~S. 2001, \apj, 551, 160

\bibitem[{{Wayth} {et~al.}(2015){Wayth}, {Lenc}, {Bell}, {Callingham},
  {Dwarakanath}, {Franzen}, {For}, {Gaensler}, {Hancock}, {Hindson},
  {Hurley-Walker}, {Johnston-Hollitt}, {Kapinska}, {McKinley}, {Morgan},
  {Offringa}, {Procopio}, {Staveley-Smith}, {Wu}, {Zheng}, {Bernardi},
  {Bowman}, {Briggs}, {Cappallo}, {Corey}, {Deshpande}, {Emrich}, {Goeke},
  {Greenhill}, {Hazelton}, {Kaplan}, {Kasper}, {Kratzenberg}, {Lonsdale},
  {Lynch}, {McWhirter}, {Mitchell}, {Morales}, {Moragn}, {Oberoi}, {Ord},
  {Prabu}, {Rogers}, {Roshi}, {Udaya Shankar}, {Srivani}, {Subrahmanyan},
  {Tingay}, {Waterson}, {Webster}, {Whitney}, {Williams}, \&
  {Williams}}]{Wayth15}
{Wayth}, R.~B., {Lenc}, E., {Bell}, M., {et~al.} 2015, Publications of the
  Astronomical Society of Australia, in press, arXiv:1505.06041

\bibitem[{{Xu} \& {Han}(2014)}]{Xu14}
{Xu}, J., \& {Han}, J.~L. 2014, \mnras, 442, 3329

\end{thebibliography}

\end{document}